# Electromagnetically induced grating based on strongly coupled disperse red-1 molecules

Xiaotong Bu[1,2,3], Jingsong Yuan[1,2,3], Chunliang Wang[1,2,3*]

Key Laboratory of UV Light-Emitting Materials and Technology of Ministry of Education, Northeast Normal University, Changchun 130024, China

Email: wangcl493@nenu.edu.cn

**Abstract**

In this study, optically responsive exciton-polariton were generated via strong coupling between DR1 molecules and an F-P microcavity. An electromagnetically induced grating (EIG) was constructed using two-beam interference to investigate EP modulation effects. EP formation was confirmed by transmission spectroscopy, angle-dependent measurements, and k-space reflection. Rabi splitting increased with DR1 concentration, indicating enhanced light-matter interaction. Under 532 nm laser illumination, reversible Rabi splitting changes occurred: upper/lower polariton peaks shifted by 16 nm/8 nm toward the 490 nm absorption peak with a 2% transmittance increase, recovering after laser removal. Introducing EP into EIG enhanced first-order diffraction intensity by 2–10 times compared to the non-coupled state, with peak positions shifting 8–30 nm under EP regulation. Diffraction angles varied within 2°, correlating with coupling strength. Compared with the bare DR1/PMMA film, the EIG diffraction signal of the strong coupling device was 2–7 times stronger, with clearer resolution and faster optical response near the resonance position, attributed to the enhanced light-matter interaction in the EP system.

## 1. Introduction

In 1990, Harry first discovered the Electromagnetically Induced Transparency (EIT) phenomenon in atomic vapor. Through quantum interference of pump light and probe light during different energy level transitions, an opaque atomic medium can generate a transparent window in a narrow region of a broad absorption band [1]. In 1995, Hemmer et al. reported [2] that the diffraction of laser on purely coherent gratings was successfully observed experimentally in thermal sodium vapor, achieving high phase-conjugate gain and fast response, which provides possibilities for developing practical, portable, and high-gain optical signal processing devices. Relying on the EIT phenomenon [3], the Electromagnetically Induced Grating (EIG) effect was proposed by Ming Xiao's research team in 1998 [4-5], in which advanced diffraction of a weak probe field is generated through the interaction between a strong coupling standing wave field and atoms. In 2005, an electromagnetically induced absorption grating was constructed in a three-level atomic vapor system to study the reflection characteristics of the EIG probe light and its relationship with laser frequency detuning, realizing all-optical two-port signal routing and all-optical switching functions, and providing a new approach for quantum information processing and quantum networks [6]. In 2006, externally induced optical nonlinearity created periodic mod ulation in a uniform absorption medium to achieve optical tuning of photonic bandgaps. In cold atomic systems, pump light forms standing waves after reflection, and the Rabi frequency changes periodically in space to modulate the absorption and dispersion of probe light. This can be used for quantum optical storage, optical switching, research on cold-atom Bragg processes, and macroscopic entanglement demonstration, and is also expected to be extended to solid-state materials [7]. In 2017, a scheme for implementing beam splitters and beam routers through electromagnetically induced blazed gratings was proposed in a four-level double-Λ system driven by intensity-modulated coupling fields and incoherent pump fields [8]. Due to its many advantages, EIG has attracted widespread attention and has been applied in research fields of optical quantum devices such as optical signal storage and reading, all-optical switching and routing, coherently induced photonic bandgaps, quantum Talbot effects, and beam splitters [9-11]. However, the realization of traditional EIG [12-16] relies on cold atomic ensembles or room-temperature atomic cells, leading to large system volume and strict environmental stability requirements, which severely limit its applications in integrated photonics and scalable quantum devices. Therefore, implementing EIG in solid-state nanoscale thin-film systems can seamlessly interface with many existing nano-optical systems. In this thesis, light-responsive exciton-polariton (EP) were prepared via strong coupling of Disperse Red 1 (DR1) [17-21] and Fabry-Pérot microcavities (F-P) [22-24]. On this basis, EIG was formed by irradiating interference fringes from two light beams. The introduction of EP makes the overall diffraction signal of EIG clearer, with the diffraction signal intensity increased by 2-10 times compared to that without EP. The position of the diffraction peak shifts by 8-30 nm under different EP modulations, and the diffraction angle corresponding to the strongest diffraction signal also shifts within 2°. This innovative combination expands the application scope of strong coupling in the optoelectronic field and provides a theoretical reference and experimental basis for developing new optical devices.

## 2. Research and Discussion

In this experiment, the organic azo dye used is DR1 molecule, with a molecular weight of 314.34 g/mol. It can absorb the energy of pump light to undergo photoisomerization from the trans structure to the cis structure. The molecular structure of DR1 consists of nitrobenzene units and belongs to a monoazo compound, characterized by an N=N double bond connecting two aromatic rings, which endows it with certain conjugacy.This unique structure results in their maximum absorption peak ranging from 350 nm to 600 nm.

To achieve strong coupling between DR1 molecules and the F-P for forming EP, a silver layer with a thickness of 30 nm was magnetron-sputtered onto a 2 cm × 2 cm glass substrate. Then, a polymer film doped with polymethyl methacrylate (PMMA) with a thickness of approximately 110 nm was prepared using a spin coater. The volume ratio of DR1 to PMMA was 1:1, and the mixture was spin-coated onto the silver film at a speed of 1350 rad/min. After the sample dried, another 30 nm silver layer was sputtered on top to obtain the strong coupling device.

In this experiment, a self-built electromagnetic induction grating preparation and diffraction signal detection system was used to measure the spectral characteristics. The laser light source was expanded and split into two vertically polarized beams by a beam-splitting prism, which were adjusted by mirrors to focus on a single point. In the detection optical path, the probe light output by the light source was converted into parallel light by a collimator, passed through a polarizer, mirror, and lens, and then transmitted through the strong coupling device. The fir st-order diffraction signal was detected by a spectrometer. As shown in Figure 1, the yellow line represents the probe optical path using a halogen-tungsten lamp as the visible light source, and the green line represents the excitation optical path using a 532 nm (single longitudinal mode) laser.

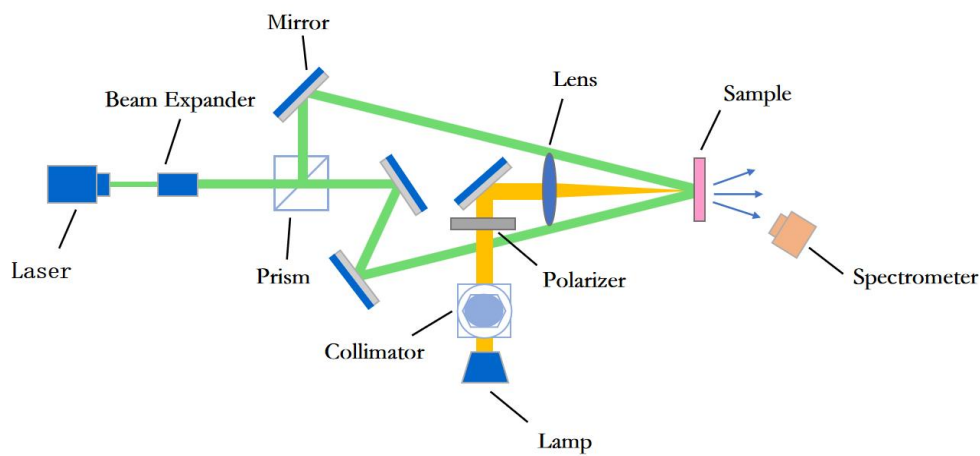

Figure1.Schematic diagram of the experimental setup for preparing electromagnetically indeced grating (EIG) via two-beam interference and detecting their diffraction signals.

## 2.1 Formation of EP

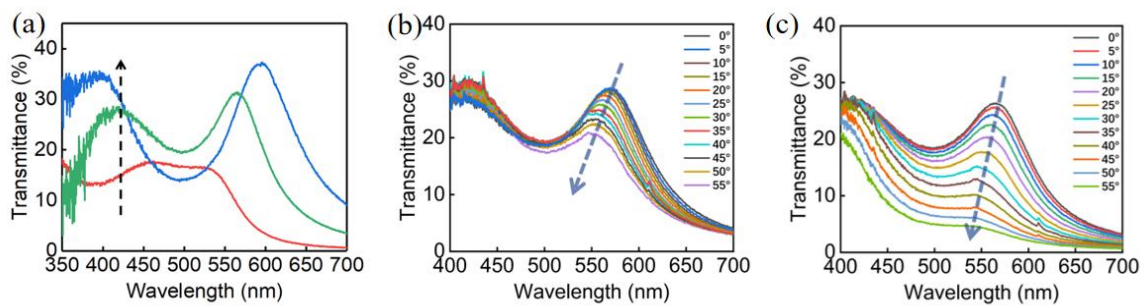

Figure 2. (a) Transmission spectra of the strong coupling system showing increased Rabi splitting with increasing DR1 concentration; (b),(c) Variable-angle transmission spectra of the strong coupling device under P-light and S-light excitation, respectively, upon direct incidence from the air side.

The prepared DR1/PMMA mixed solution is placed in an F-P cavity formed by two silver films. By adjusting the thickness of the film to control the length of the optical cavity, the resonant mode of the cavity is matched with the molecular energy levels, enabling the system to enter a strong coupling state and form EP. Figure 2(a) shows the transmission spectrum of the strong coupling device, where obvious Rabi splitting can be observed. As the concentration of DR1 increases, the Rabi splitting value becomes larger, and the dip at 490 nm gradually deepens. In the strong coupling system, DR1 molecules interact with photons. When the molecular concentration of the sample increases, the number of molecules participating in coupling per unit volume increases. The Rabi frequency is related to the coupling strength $g$ between molecules and photons. According to the theoretical model, within a certain range, as the DR1 molecular concentration increases, $g$ increases, and the Rabi splitting also increases. Figures 2(b) and 2(c) show that after changing different incident angles, both P-light and S-light are split into two peaks near 490 nm, indicating that strong coupling between DR1 and the F-P mode occurs to generate EP. With the change of the incident angle, it is found that the positions of the two peaks shift blue as the incident angle increases, and the transmittance shows an obvious decreasing trend.

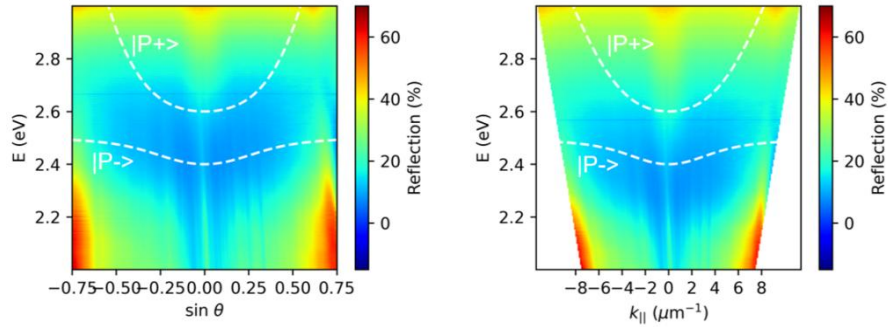

Figure 3.Reflection spectrum of the Ag-DR1/PMMA-Ag structure.

Figure 3 shows the wavevector-energy dispersion diagram of the strong coupling device obtained by the self-assembled k-space reflection spectroscopy testing system in the experiment. In the Ag-DR1/PMMA-Ag structure, the reflection spectral information varies at different incident angles, where the vertical axis is energy $E$, and the horizontal axis $k_{//}$ is the wavevector component within the plane of the device. The color from blue to red indicates that the reflectivity gradually increases. When the sample is in the strong coupling state, new states |P+> and |P-> are formed. At the resonant wavelength, both the transmittance and absorption increase, while the reflectivity decreases. In the dispersion diagram, two curved regions can be observed at the resonant wavelength, which are axisymmetric about $k_{//}=0$. The white dashed lines correspond to the positions of the upper polariton and lower polariton, and the intermediate region does not intersect. In the dispersion diagram, the blue part has lower reflectivity, which is caused by energy transfer at this position after EP formation, leading to reduced reflectivity at the resonant position. The white dashed lines in the figure represent the theoretical calculation results of |P+> and |P->, and the pseudocolor plot fully displays the trend of k-space reflection, indicating the direction of the anti-crossing signal. We have confirmed the strong coupling state of the system and the formation of EP from three aspects: transmission spectra, variable-angle strong coupling tests, and k-space reflection spectroscopy tests.

## 2.2 Optically responsive exciton-polariton

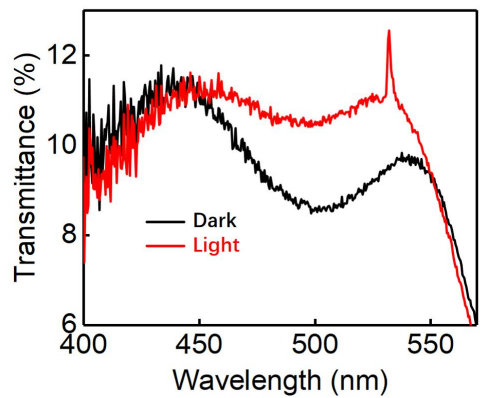

Figure 4. All-optical modulation curves of the strong coupling device by a 532 nm laser.

Azobenzene-based molecules can exhibit significant photo-responsive properties under linearly polarized light irradiation. The molecular electric dipole moment has a component along the polarization direction of the pump light. When azobenzene molecules undergo cis-trans isomerization, their molecular orientation is also rearranged. After stabilization, the direction of the molecular electric dipole moment becomes perpendicular to the polarization direction of the pump light. This rearrangement of molecular orientation imparts anisotropy to these organic molecules, known as photoinduced anisotropy.

Figure 4 shows that after the device enters the strong coupling state to form EP, the transmission spectrum of the system changes significantly under photoexcitation. Upon light illumination, the Rabi splitting of the strong coupling system decreases. The peaks of |P+> and |P-> shift toward the absorption peak (490 nm) of DR1 molecules by 16 nm and 8 nm, respectively, and the transmittance increases by approximately 2%. This occurs because DR1 molecules undergo both photoinduced isomerization and photoinduced anisotropy after absorbing pump light energy, leading to a decrease in the absorbance of the device along the polarization direction of the pump light. As Rabi splitting is a function of absorbance and is positively correlated with the square root of the film's absorbance, the Rabi splitting decreases accordingly. At this time, the transmittance near 500 nm increases due to the closer proximity of |P+> and |P->. This is highly conducive to studying the properties of EP generated when excitons couple with the vacuum field to different degrees. When the 532 nm laser is turned off, the transmittance of the device gradually decreases, and the spacing between |P+> and |P-> also recovers.

## 2.3 EIG Modulated by EP

When the energy exchange rate between excitons and photons in a confined space exceeds the dissipation rate of the system, the system enters a strong coupling state, thereby forming a new quasiparticle EP. In the experiment, two interference fringes irradiate the strong coupling device, and the two coherent wavefunctions are respectively: $E_1 = E_{01}\cos(\omega t - k_1 r_1 + \varphi_1)$

$$E_2 = E_{02}\cos(\omega t - k_2 r_2 + \varphi_2)$$

The combined light intensity *I* after superposition of the two beams is:

$$I = I_1 + I_2 + 2\sqrt{I_1 I_2}\cos\Delta\varphi$$

Therefore, the grating equation (expression for the grating constant) formed by two-beam interference is:

$$d = \frac{\lambda}{2\sin\theta}$$

Where *d* is the grating constant of the equivalent grating formed by the interference fringes, *λ* is the wavelength of the incident light, and *θ* is half of the angle between the propagation directions of the two beams. This equation describes the relationship between the fringe spacing (grating constant) formed by two-beam interference, the incident light wavelength, and the angle between the two beams.

When two coherent light beams irradiate a medium containing azo dyes, bright and dark interference fringes are formed. In the bright fringe regions, azo dye molecules absorb photon energy and undergo photoisomerization, transforming from the trans structure to the cis structure. The cis molecules differ in shape and dipole moment from the trans molecules and will reorient under the action of the light field. The refractive index of the medium is closely related to its polarizability, which in turn depends on the orientation and distribution of the molecules. When the molecular orientation undergoes spatial modulation, the polarizability of the medium also changes periodically, thereby causing periodic changes in the refractive index. Assuming the relative permittivity of the medium is *ε*, the refractive index *n*=*ε*. During the modulation of molecular orientation, the relative permittivity periodically changes with the molecular orientation, thus forming a structure with periodic refractive index variations in the medium.

To illustrate the influence of EP on EIG, we prepared two strong coupling devices by adjusting the thickness of the DR1/PMMA polymer layer film, and another device where the optical mode of the resonant cavity did not match the exciton energy levels, resulting in a device that did not enter the strong coupling state (zero coupling). The transmission spectra are shown in Figure 5(a).

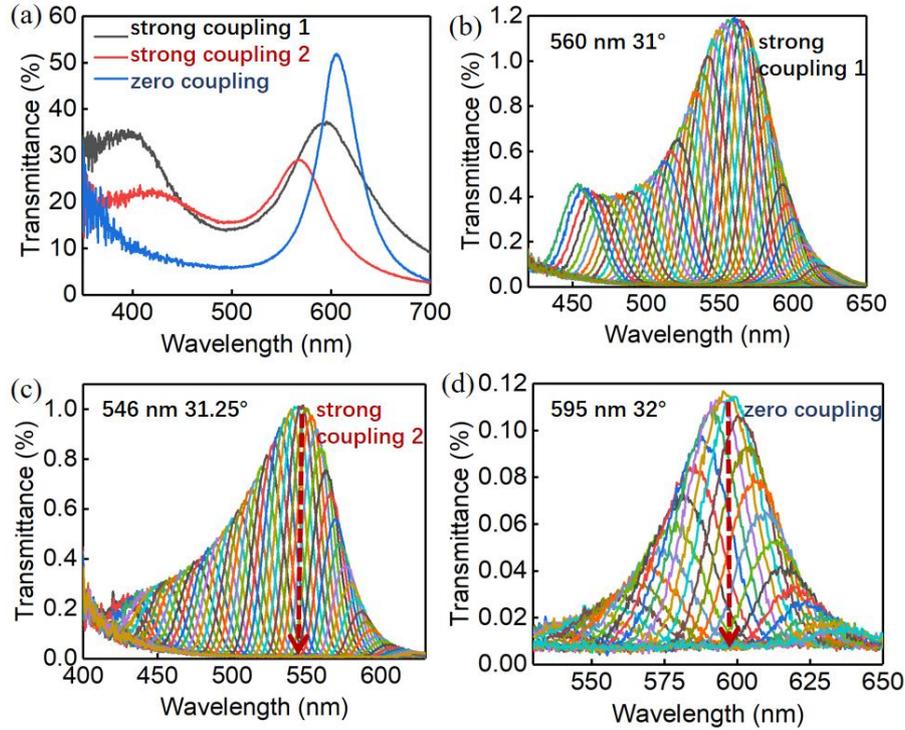

Figure 5. (a) Transmission spectra of three samples; (b, c, d) Diffraction signals corresponding to the three coupling samples.

From the figure, we observe that for the strong coupling 1 sample, the |P+> and |P-> states are located at 400 nm and 592 nm, respectively; for the strong coupling 2 sample, |P+> and |P-> are at 415 nm and 566 nm, respectively. In contrast, the zero-coupling sample only exhibits one transmission peak |P-> at 603 nm. Due to the significant mismatch between the optical mode of the resonant cavity and the exciton energy levels, the density of states for |P+> is nearly zero, so |P+> is not observable in the transmission spectrum, and EP is not formed.

The above three samples with different EP states were respectively placed on a sample stage for uniform EIG construction. Using a halogen-tungsten lamp for detection and a spectrometer to receive signals, the obtained spectra are shown in Figures 5(b), (c), and (d). Introducing EP into EIG enhances its first-order diffraction signal, with the detected diffraction signal wavelength range spanning 400 nm to 650 nm. First, in the zero-coupling sample, without the influence of EP coupling, the detected diffraction signal is weak. In contrast, the diffraction signal intensity in the strong coupling samples (including strong coupling 1 and 2) is approximately 10 times higher than that of the zero-coupling sample. Second, when EP is introduced into EIG, the peak wavelengths of the diffraction peaks undergo blue shifts of varying degrees. For the strong coupling 1 sample, a double-peak structure appears in the diffraction signal: the |P-> state is located at 592 nm in the transmission spectrum, but the diffraction peak maximum occurs at 560 nm with a diffraction angle of 31°, resulting in a 32 nm deviation in the diffraction peak position. For the strong coupling 2 sample, the Rabi splitting energy is slightly smaller, with fewer molecules undergoing strong coupling, and only a tendency toward a double-peak structure is observed in the diffraction signal spectrum: the |P-> state is at 566 nm in the transmission spectrum, the diffraction peak maximum occurs at 546 nm with a diffraction angle of 31.25°, and the diffraction peak position shifts by 20 nm. In the zero-coupling sample, the |P-> state is at 603 nm in the transmission spectrum, the diffraction peak maximum occurs at 595 nm, and the diffraction angle is 32°.

In summary, after introducing EP into EIG, both the diffraction signal intensity and the degree of diffraction peak position shift increase with the enhancement of the coupling strength.

## 2.4 Electromagnetically indeced grating in DR1/PMMA Polymer Films

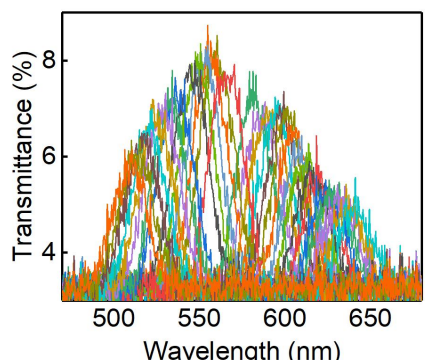

Figure 6. First-order diffraction signal of EIG formed by the bare film.

The prepared DR1/PMMA mixed solution is placed on a glass substrate, and a DR1/PMMA polymer film with a thickness of approximately 110 nm is spin-coated using a spin coater to obtain a bare film device. When two beams of 532 nm coherent light irradiate, interference fringes are formed. Figure 6 shows the first-order diffraction signal of EIG formed by the bare film, with a detectable wavelength range of 500 nm-650 nm and a weak diffraction signal.

Bare film and strong coupling devices were selected for transmission spectrum testing. The bare film exhibits absorption at 490 nm. In the two coupling devices, the short-wavelength resonant |P+> and |P-> are located at 395 nm and 575 nm, while the long-wavelength resonant |P+> and |P-> are at 405 nm and 600 nm. As shown in Figure 7, the samples were subjected to 15 minutes of two-beam interference light irradiation to form EIG, and diffraction signal detection was performed at 490 nm, 557 nm, 590 nm, and 620 nm with light on and off.

Figures (a) and (b) show the detection at 490 nm and 557 nm. Due to the strong 532 nm writing light signal and the weak diffraction signal of the bare film, the bare film diffraction signal is low and undetectable in figures (a) and (b). In figures (c) and (d), the detection wavelengths are 590 nm and 620 nm. The first-order diffraction signal of EIG constructed by the bare film is weak, while the diffraction signal of the strong coupling device is 2-7 times higher than that of the bare film device. At this time, the detection wavelength is closer to the long-wavelength resonance position. When the writing light is immediately turned off, the diffraction signal of the long-wavelength resonance sample rapidly decreases at 590 nm and 620 nm. In contrast, the diffraction signal of the short-wavelength resonance sample decreases slowly, indicating that under the regulation of EP, excitons in the cavity respond faster to light near the resonance position.

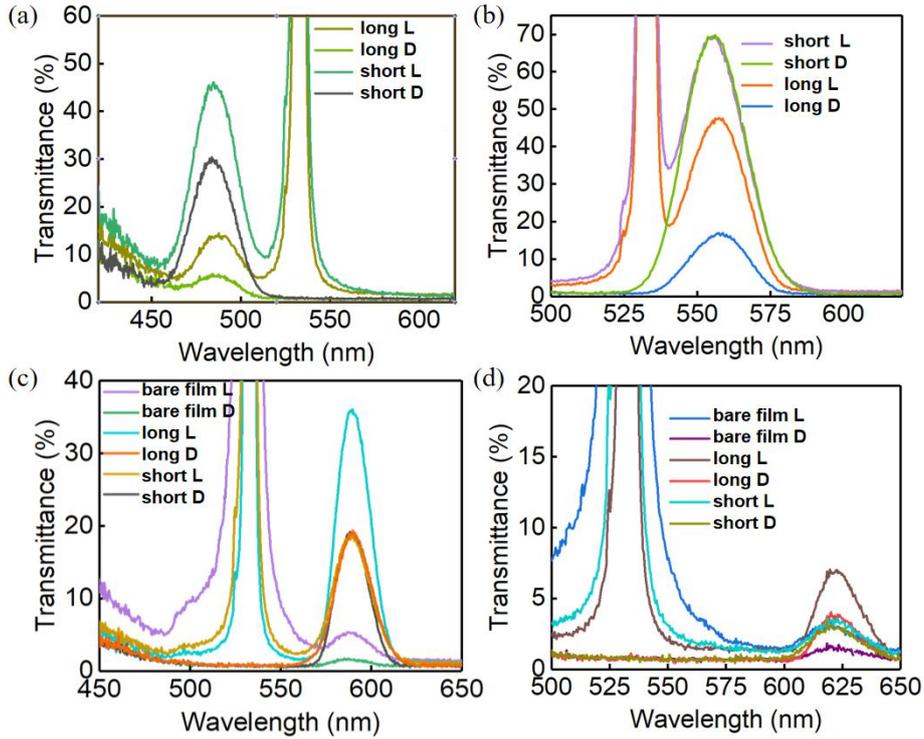

Figure 7. Detection of diffraction signals at 490 nm, 557 nm, 590 nm, and 620 nm for the DR1 bare film and strong coupling devices.

## 3. Conclusion

With the development of nanophotonics and quantum science, strong coupling, as a key direction in fundamental research, has demonstrated enormous potential in quantum optics and quantum information processing. Introducing EP into EIG and leveraging its unique half-light/half-matter properties provide new insights for the research and development of novel optical components and systems.

This paper prepares light-responsive EP based on the azo dye molecule DR1 and innovatively proposes and investigates EIG based on light-responsive EP. First, the formation of EP is confirmed by transmission spectra, variable-angle testing, and a k-space reflection spectroscopy testing system. Changing the DR1 concentration alters the minimum Rabi splitting value accordingly. Second, light-responsive EP is prepared based on the photoresponse of DR1. The regulation of EP is successfully achieved using a 532 nm laser: as the absorbance of the device parallel to the pump light polarization direction decreases, the positions of |P+> and |P-> approach the absorption peak, and they recover after the light is turned off. Finally, we construct EIG with light-responsive EP. After light-induced response, the EIG diffraction signal intensity is approximately 10 times higher than that of zero-coupling samples, and the degree of diffraction peak shift increases with the enhancement of the coupling strength. Compared with the EIG diffraction signal constructed by the DR1/PMMA bare film, the diffraction signal of the strong coupling device is 2–7 times stronger and more clearly resolved. Meanwhile, under the regulation of EP, excitons in the cavity exhibit faster photoresponses near the resonance position.

## 4.References

[1]Boller K J, Imamoğlu A, Harris S E. Observation of electromagnetically induced transparency[J]. Physical

Review Letters, 1991, 66(20): 2593.

[2] Hemmer P R, Katz D P, Donoghue J, et al. Efficient low-intensity optical phase conjugation based on coherent population trapping in sodium[J]. Optics letters, 1995, 20(9): 982-984.

[3] Fleischhauer, Michael, and Mikhail D. Lukin. "Dark-state polaritons in electromagnetically induced transparency." Physical review letters 84.22 (2000): 5094.

[4] Brown, Andy W., and Ming Xiao. "Frequency detuning and power dependence of reflection from an electromagnetically induced absorption grating." Journal of Modern Optics 52.16 (2005): 2365-2371.

[5] Brown, Andy W., and Ming Xiao. "All-optical switching and routing based on an electromagnetically induced absorption grating." Optics letters 30.7 (2005): 699-701.

[6] Brown A W, Xiao M. All-optical switching and routing based on an electromagnetically induced absorption grating[J]. Optics letters, 2005, 30(7): 699-701.

[7] Artoni M, La Rocca G C. Optically tunable photonic stop bands in homogeneous absorbing media[J]. Physical review letters, 2006, 96(7): 073905.

[8] Chen Y Y, Liu Z Z, Wan R G. Beam splitter and router via an incoherent pump-assisted electromagnetically induced blazed grating[J]. Applied Optics, 2017, 56(20): 5736-5744.

[9] Qiu T, Yang G, Bian Q. Electromagnetically induced second-order Talbot effect[J]. Europhysics Letters, 2013, 101(4): 44004.

[10] Moiseev S A, Kröll S. Complete reconstruction of the quantum state of a single-photon wave packet absorbed by a Doppler-broadened transition[J]. Physical review letters, 2001, 87(17): 173601.

[11] Deng J, Lim W F, Chen J, et al. Electromagnetically induced transparency like behavior circular dichroism effect induced by graphene grating[J]. Optik, 2022, 251: 168351.

[12] Chen Y C, Lin C W, Ite A Y. Roles of degenerate Zeeman levels in electromagnetically induced transparency[J]. Physical Review A, 2000, 61(5): 053805.

[13] Fleischhauer M, Imamoglu A, Marangos J P. Electromagnetically induced transparency: Optics in coherent media[J]. Reviews of modern physics, 2005, 77(2): 633-673.

[14] Zhang Y, Brown A W, Xiao M. Opening Four-Wave Mixing and Six-Wave Mixing Channels<? format?> via Dual Electromagnetically Induced Transparency Windows[J]. Physical review letters, 2007, 99(12): 123603.

[15] Wu Y, Yang X. Electromagnetically induced transparency in V-, Λ-, and cascade-type schemes beyond steady-state analysis[J]. Physical Review A—Atomic, Molecular, and Optical Physics, 2005, 71(5): 053806.

[16] Gu Y, Wang L, Wang K, et al. Coherent population trapping and electromagnetically induced transparency in a five-level M-type atom[J]. Journal of Physics B: Atomic, Molecular and Optical Physics, 2005, 39(3): 463.

[17] Gong, C. B., Lam, M. H.-W. & Yu, H. X. The Fabrication of a Photoresponsive Molecularly Imprinted Polymer for the Photoregulated Uptake and Release of Caffeine. Adv. Funct. Mater. 16, 1759–1767 (2006).

[18] Sekkat, Z., Morichère, D., Dumont, M., Loucif-Saïbi, R. & Delaire, J. A. Photoisomerization of azobenzene derivatives in polymeric thin films. J. Appl. Phys. 71, 1543–1545 (1992).

[19] Shi, Y., Steier, W. H., Yu, L., Chen, M. & Dalton, L. R. Large stable photoinduced refractive index change in a nonlinear optical polyester polymer with disperse red side groups. Appl. Phys. Lett. 58 1131–1133 (1991).

[20] Jaafar, A. H., Gray, R. J., Verrelli, E. & Kemp, N. T. Reversible optical switching memristors with tunable STDP synaptic plasticity: a route to Cite this: Nanoscale, 2017, 9, 17091 hierarchical control in artificial intelligent systems. (2017).

[21] Sekkat, Z. & Dumont, M. Photoassisted poling of azo dye doped polymeric films at room temperature. Appl. Phys. B Photophysics Laser Chem. 54, 486–489 (1992).

[22] Kasprzak J, Richard M, Kundermann S, et al. Bose–Einstein condensation of exciton polaritons[J]. Nature, 2006, 443(7110): 409-414.

[23]Franklin, Rhonda R., et al. "Fabry-Perot cavity antenna system having a frequency selective surface." U.S. Patent No. 10,777,901. 15 Sep. 2020.

[24]Zomer, Fabian, and Richard Cizeron. "Amplifying optical cavity of the FABRY-PEROT type." U.S. Patent No. 8,526,572. 3 Sep. 2013.